\documentclass{ws-jai}
\usepackage[flushleft]{threeparttable}
\usepackage[breaklinks,colorlinks, urlcolor=blue,citecolor=blue,linkcolor=blue]{hyperref}
\usepackage{graphicx}
\usepackage{color}
\usepackage{amsmath}
\usepackage{color}
\usepackage[bottom]{footmisc}
\usepackage{adjustbox}
\usepackage{blindtext}
\usepackage{amsfonts}
\usepackage{amssymb}
\usepackage{amsmath}

\def\prizm{PRI$^{\rm Z}$M}
\def\scihi{SCI-HI}
\begin{document}

\catchline{}{}{}{}{} 

\markboth{Liju Philip et al.}{Probing Radio Intensity at high-Z from Marion: 2017 Instrument}

\title{Probing Radio Intensity at high-Z from Marion: 2017 Instrument}

\author{L. Philip$^{1}$, Z.~Abdurashidova$^{2}$, H.~C.~Chiang$^{3,4}$, N.~Ghazi$^{3}$, A.~Gumba$^{1}$, H.~M.~Heilgendorff$^{3}$, J.~Hickish$^{2}$,  J.~M.~J{\'a}uregui-Garc{\'i}a $^{5}$, K.~Malepe$^{6}$, C.~D.~Nunhokee$^{7}$, J.~Peterson$^{5}$, J.~L.~Sievers$^{1,4}$, V.~Simes$^{3}$, and R.~Spann$^{8}$}

\address{
$^{1}$School of Chemistry and Physics, University of KwaZulu--Natal, Durban 4000, South Africa\\
$^{2}$Department of Astronomy, University of California, Berkeley, California 94720, USA\\
$^{3}$School of Mathematics, Statistics, and Computer Science, University of KwaZulu--Natal, Durban 4000, South Africa\\
$^{4}$National Institute for Theoretical Physics, Durban 4000, South Africa\\
$^{5}$Department of Physics, Carnegie Mellon University, Pittsburgh, PA 15213, USA\\
$^{6}$South African National Space Agency, Hermanus 7200, South Africa\\
$^{7}$Department of Physics and Electronics, Rhodes University, Grahamstown 6140, South Africa\\
$^{8}$South African Square Kilometre Array, Pinelands 7405, South Africa
}

\maketitle

\begin{history}
\received{(to be inserted by publisher)};
\revised{(to be inserted by publisher)};
\accepted{(to be inserted by publisher)};
\end{history}

\begin{abstract}
We introduce Probing Radio Intensity at high-Z from Marion (\prizm), a
new experiment designed to measure the globally averaged sky
brightness, including the expected redshifted 21~cm neutral hydrogen
absorption feature arising from the formation of the first stars.
\prizm\ consists of two dual-polarization antennas operating at
central frequencies of 70 and 100~MHz, and the experiment is located
on Marion Island in the sub-Antarctic.  We describe the initial design
and configuration of the \prizm\ instrument that was installed in
2017, and we present preliminary data that demonstrate that Marion
Island offers an exceptionally clean observing environment, with
essentially no visible contamination within the FM band.
\end{abstract}


\section{Introduction}\label{sec:intro}

The 21-cm line of neutral hydrogen provides one of the best ways to
search for cosmic dawn, the epoch of the first luminous objects
\citep{Furlanetto2006, Pritchard2012}.  Cosmic dawn is expected to
have imprinted a characteristic absorption feature in the intensity spectrum of
the globally (all-sky) averaged brightness, with
a typical amplitude of $\sim200$~mK. This feature contains a wealth of
information about these first objects ({\textit{e.g.}}
\citet{Cohen2017}).  The recent report of a 500~mK deep
absorption feature centered at 78~MHz \citep{Bowman2018} substantially exceeds
the expected amplitude. If confirmed, this strong absorption may require
physical processes well beyond those included in previous models,
including early dark matter interactions \citep{Barkana2018} and extra
radiation at low frequencies \citep{Pospelov2018,Dowell2018}.

The detection of the cosmic dawn absorption feature is made
challenging by bright astrophysical foregrounds, ionospheric
fluctuations, radio-frequency interference (RFI), and instrumental
systematic errors.  The Milky Way is $10^3$--$10^4$ times brighter
than the expected cosmological signal within the frequency range of
70--100~MHz.  Hence, systematic errors arising from factors like
imperfect Galaxy modelling or frequency structure in the instrument
response can mimic the cosmic dawn signal.  The ionosphere is
time-varying and both absorbs and refracts at these frequencies, thus
introducing additional measurement noise and uncertainty.
Furthermore, time-invariant broadband RFI is indistinguishable from a
cosmic dawn signal, pushing experiments to some of the most remote
locations on Earth.  Interference from FM radio lies in a frequency
range of $\sim$88--105~MHz and extends hundreds of kilometers from the
source stations, presenting a serious challenge for global signal
observations~\citep{Voytek2014, Monsalve2017}.  Because of the
measurement difficulty from most continental sites, there have been
some proposals to go to the far side of the moon ({\textit{e.g.}}
DARE, \citet{Burns2017, Burns2012}).

\begin{figure}
	\centering
	\includegraphics[width=0.75\textwidth]{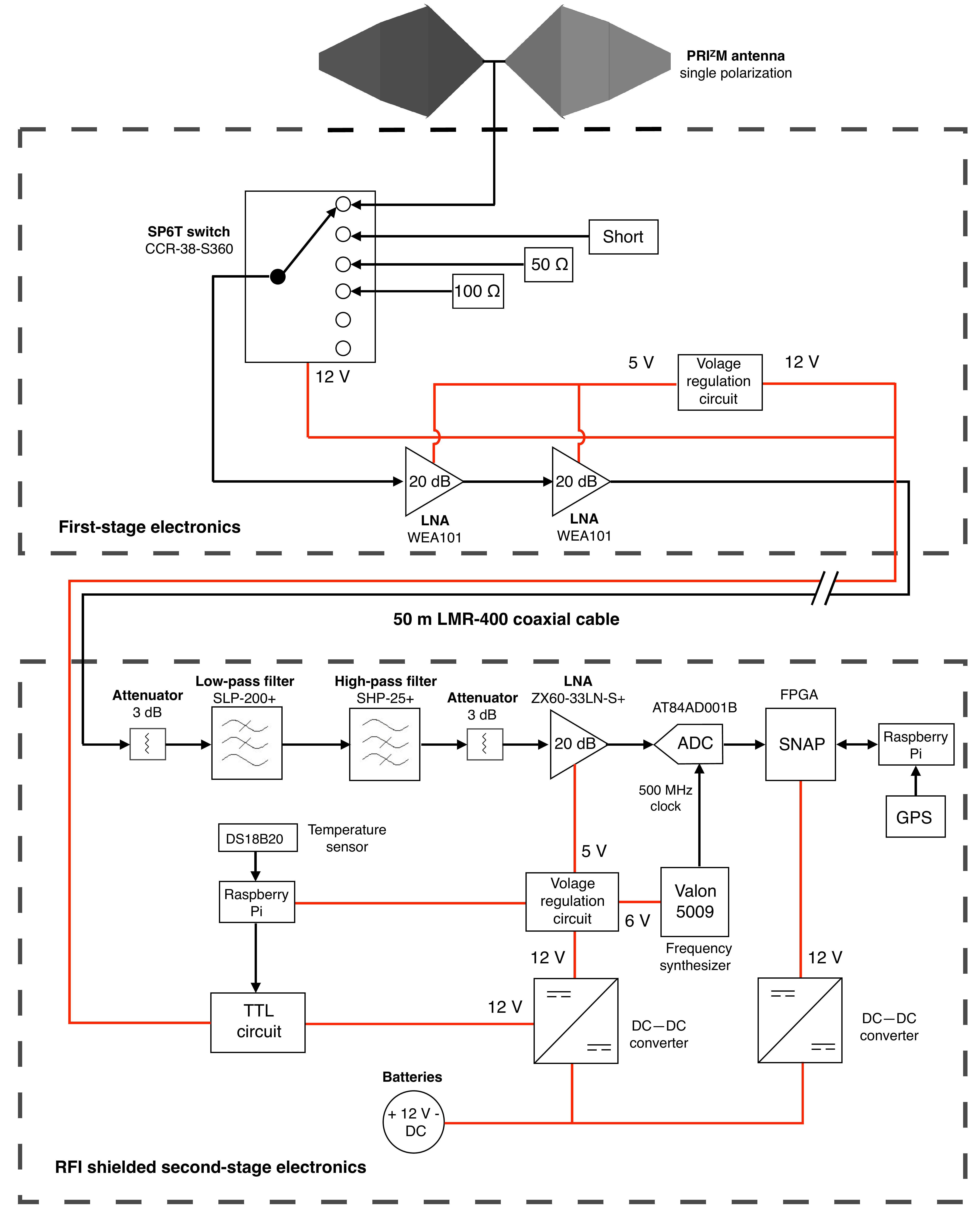}
	\caption{Block diagram for a single polarization
		\prizm\ antenna. The upper and lower dashed boxes represent
		the first and second stages of the electronics chain. The
		two stages are separated by 50~m to reduce contamination
		from self-generated RFI.}
	\label{fig:bd}
\end{figure}

Here we describe Probing Radio Intensity at high-Z from Marion
(\prizm), a new experiment that is designed to search for the cosmic
dawn absorption feature in globally averaged redshifted 21-cm
emission.  \prizm\ joins several other past and current global signal
experiments, including the Experiment to Detect the Global EoR
Signature \citep[EDGES,][]{Monsalve2017, Mozden2017}, the
Large-aperture Experiment to detect the Dark Age
\citep[LEDA,][]{Price2017, Bernardi2015}, Shaped Antenna measurement
of background Radio Spectrum \citep[SARAS,][] {Singh2017, Patra2013},
Broadband Instrument for Global Hydrogen Reionisation Signal
\citep[BIGHORNS,][]{Sokolowski2015}, and Sonda Cosmol{\'o}gica de las
Islas para la Detecci{\'o}n de Hidr{\'o}geno Neutro
\citep[SCI-HI,][]{Voytek2014}.  \prizm\ is set apart by its location
on Marion Island, which, at 2000~km from the nearest permanent
inhabitants, is one of the most remote locations on Earth, allowing
RFI-free access to the full frequency range of the global signal.

\section{Instrument overview}

The \prizm\ experiment comprises two dual-polarization radiometers
operating at center frequencies of 70 and 100~MHz; the combined
frequency coverage of the two radiometers spans 30--200~MHz.
Figure~\ref{fig:bd} shows a block diagram of the \prizm\ signal chain
for a single polarization.

\subsection{Antenna} \label{sec:ant}

\begin{figure}
	\centering
	\includegraphics[width=0.55\textwidth]{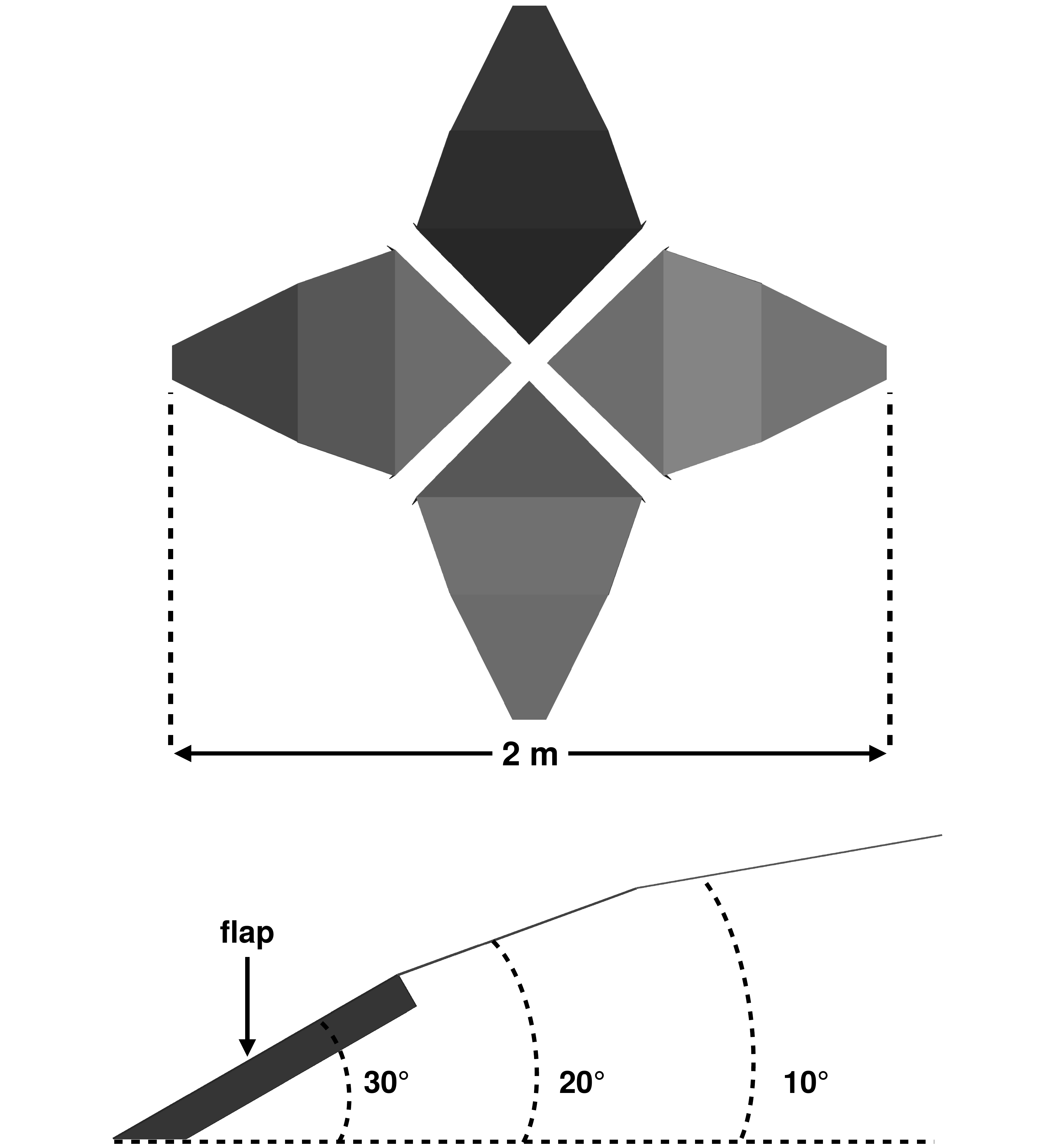}
	\caption{\textit{Top:}~\prizm\ has a modified four-square
          antenna design that consists of two crossed dipoles, which
          are aligned north--south and east--west.  The 100~MHz
          assembly is roughly 2~m $\times$ 2~m. \textit{Bottom:}~Side
          view of a single antenna petal.  The innermost triangular
          section has flaps on both sides that are bent downward with
          an opening angle of 106$^\circ$.} \label{fig:petal}
\end{figure}

\begin{figure}
	\centering
	\includegraphics[width=0.8\textwidth]{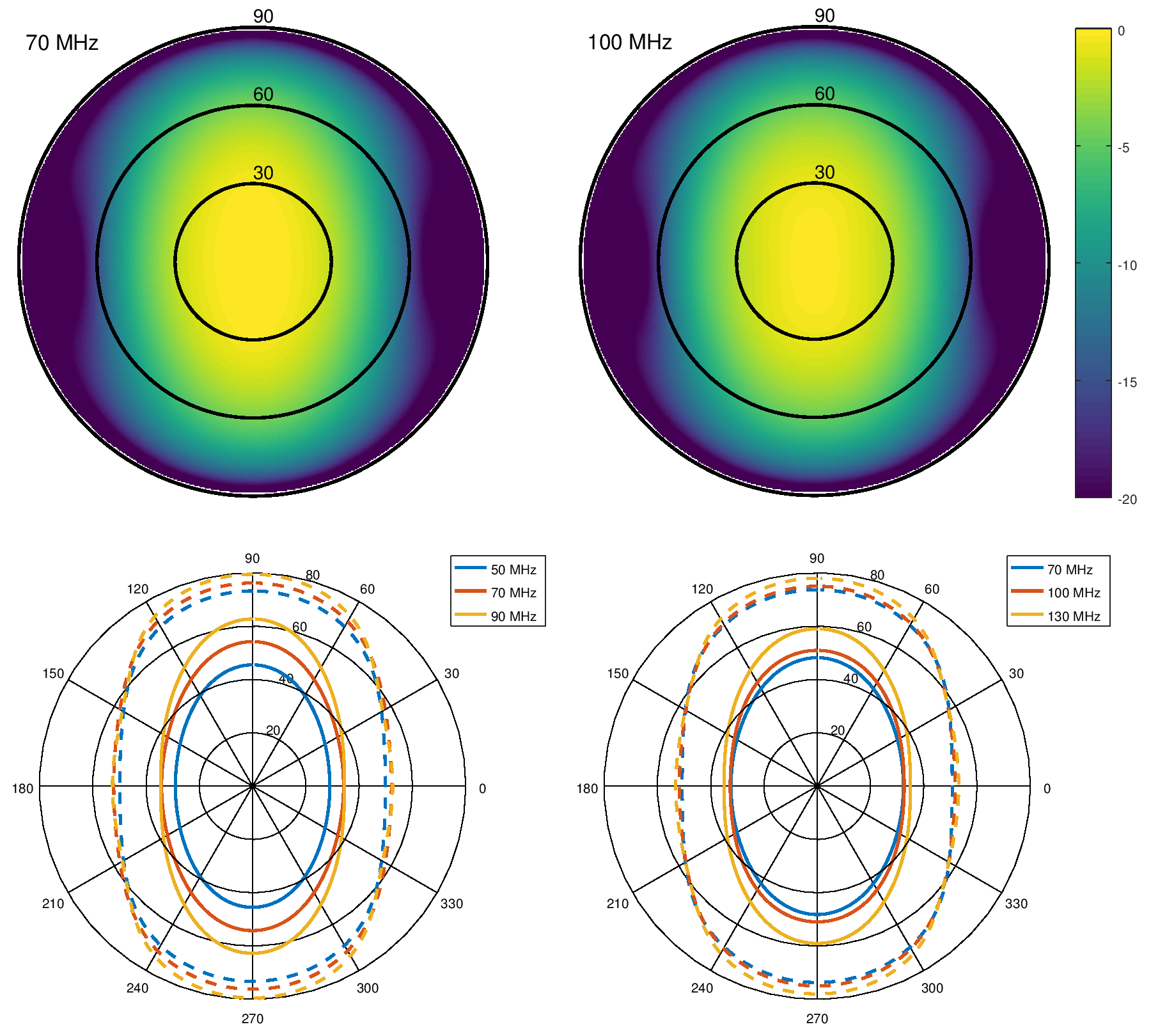}
	\caption{FEKO simulated \prizm\ beam responses for a single
          polarization of the 70~MHz (left) and 100~MHz (right)
          antennas.  The top row shows the logarithmic beam amplitude
          relative to the peak, calculated at the central frequency,
          and the bottom row illustrates how the $-3$~dB (solid lines)
          and $-10$~dB (dashed lines) contours vary with frequency.}
	\label{fig:beam}
\end{figure}

The \prizm\ radiometers use the HIBiscus four-square antenna
design~\citep{Jauregui2017} that was originally developed for the
\scihi\ experiment.  As shown in Figure~\ref{fig:petal}, each antenna
consists of four petals that form a pair of tightly-coupled crossed
dipoles.  Each petal has three angled trapezoidal facets and is made
of aluminum with black powder coating, which absorbs optical and
infrared solar radiation, increasing sublimation of snow and ice.  The
70~MHz and 100~MHz antenna assemblies are roughly 3~m and 2~m on a
side, respectively.  The physical scales yield optimized performances
in frequency ranges of 50--90~MHz for the 70~MHz antenna and
70--130~MHz for the 100-MHz antenna.  The antennas are stationary and
point at zenith, and the polarization axes are aligned with the
cardinal directions.

The antenna petals are supported by a fiberglass structure that
consists of a central column, angled trusses, and a rigid grate that
forms the base.  The central column houses the first stage
electronics, the trusses provide the necessary rigidity for sustaining
high winds, and the base provides both anchoring weight and a flat
mechanical reference surface. The entire antenna and fiberglass
structure sits on top of a ground screen that is roughly 10-m on a
side and is made of welded wire mesh.  The ground screen is further
extended by 16 10-m wire spokes that are attached to the perimeter and
run radially outward.  Electromagnetic simulations of the ground
screen, including the damp soil at the Marion deployment site, suggest
that these radial extensions have the same effect as extending the
area of the ground screen; however, the radial extensions are
significantly easier to install.  The combination of the mesh sheet
and radial extensions is intended to minimize ground-plane resonances
that create weak ripples in the spectrum of the reflected power and
the antenna beam pattern.  Figure~\ref{fig:beam} shows the predicted
beam patterns of the antennas, as simulated by
FEKO~\footnote{https://altairhyperworks.com/product/FEKO}.

\subsection{First stage electronics} \label{sec:fse}

\begin{figure}
	\centering
	\includegraphics[width = 0.6\textwidth]{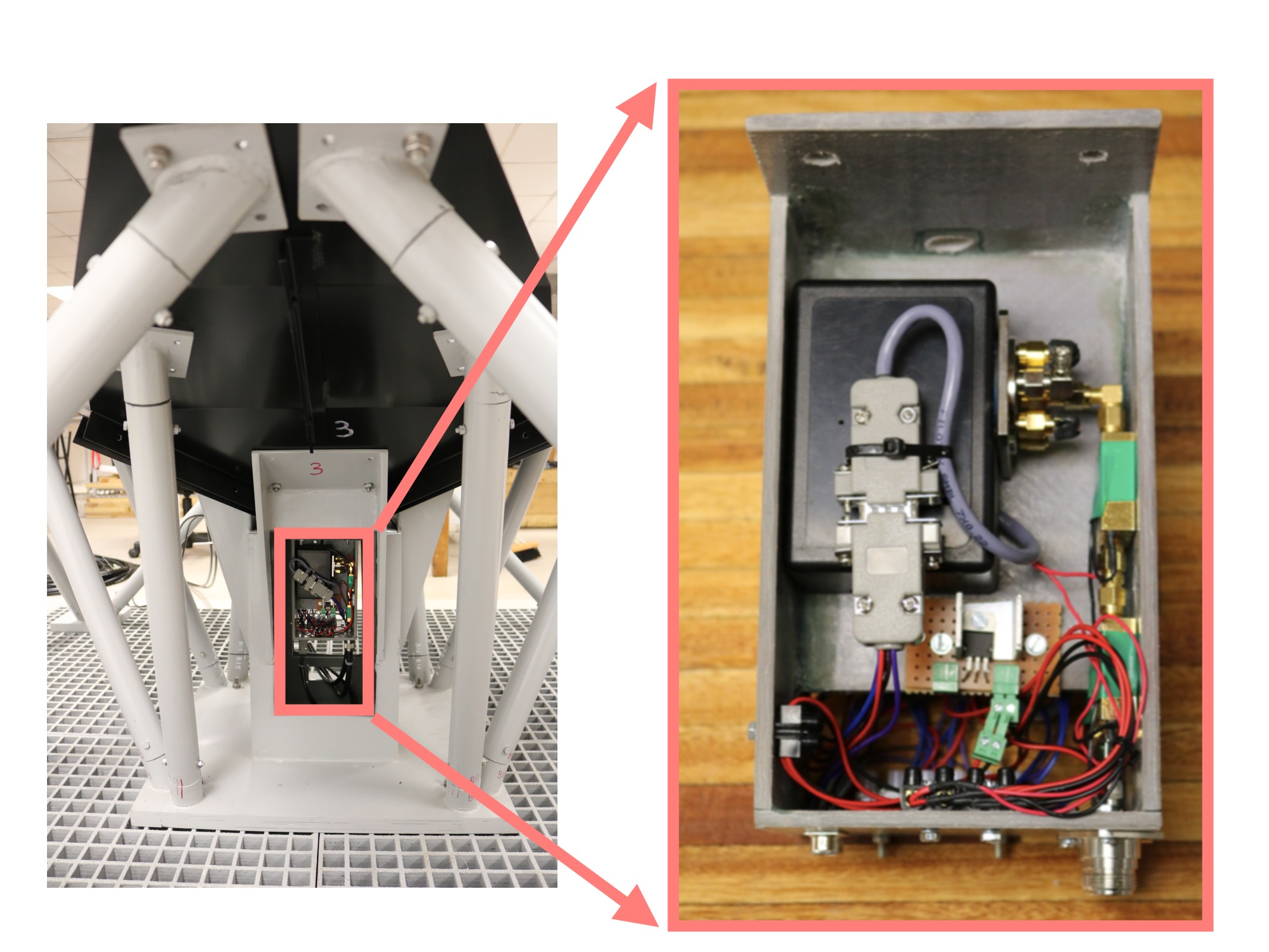}
	\caption{\textit{Left}: The first stage electronics box is located
		directly underneath the antenna petals and is visible through one
		of the windows in the central column. \textit{Right}: An
		electromechanical switch (black box) selects between the antenna,
		50$\Omega$ load, 100$\Omega$ load, and a short.  The output of the
		switch is connected to two LNAs that have a combined gain of
		40~dB.}
	\label{fig:fs}
\end{figure}

The first stage electronics are situated directly underneath the
antenna petals and are housed in a fiberglass box, as shown in
Figure~\ref{fig:fs}.  The box is vertically divided into two halves,
and each half accommodates electronics that serve one polarization.
Two windows on the central column allow access for installing and
servicing the first stage electronics box.

The first element in the electronics chain is a CCR-38S series SP6T
electromechanical switch that selects between the antenna and several
calibrator sources (50$\Omega$ terminator, 100$\Omega$ terminator, and
a short) that are installed on the switch terminals.  The output of
the switch is connected to a chain of two Richardson
RFPD~\footnote{http://www.richardsonrfpd.com/} WEA101 low noise
amplifiers (LNAs), each with a gain of 20~dB and a noise figure of
1.0~dB.  The amplifier output is routed to the second stage
electronics using direct burial LMR-400 coaxial cable that is 50~m
long, allowing the two antennas to be separated from each other, and
reducing potential contamination from self-generated RFI emissions
from our sampling system.  Each end of the cable is wrapped with a few
meters of stainless steel wire mesh cloth in order to protect the
electronics from mouse damage.

\subsection{Second stage and readout electronics}

The second stage and readout electronics are enclosed in a Faraday
cage located 50~m away from the antenna.  The second stage electronics
chain consists of a Minicircuits SLP-200+ low pass and SHP-25+ high
pass filter that band limits the RF signal to 30--200 MHz before
secondary amplification with a Minicircuits ZX60-33LN-S+ amplifier
that provides 20~dB gain. This amplified signal is then passed to the
readout electronics.

\begin{figure}
	\centering
	\includegraphics[width = 0.5\textwidth]{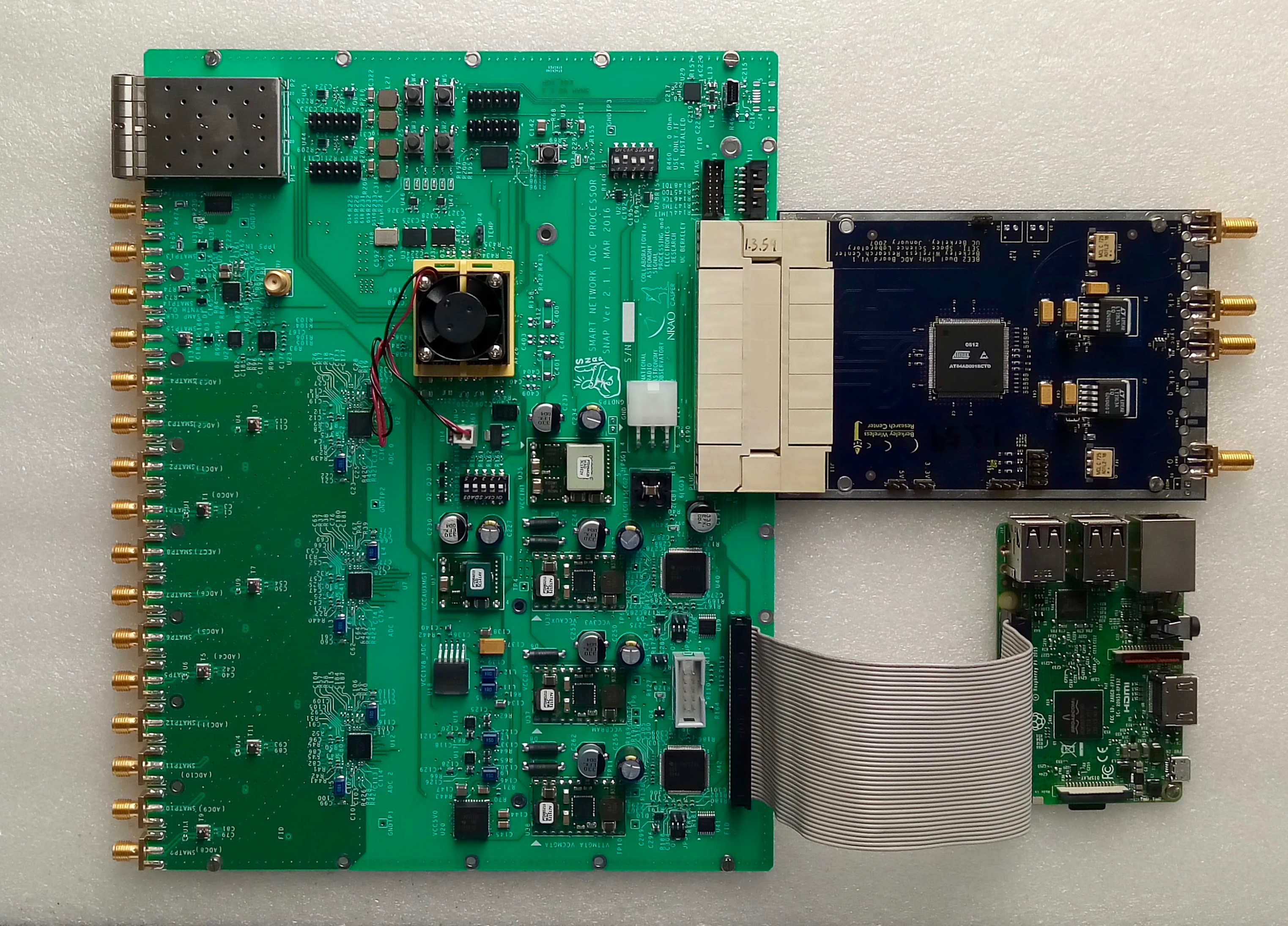}
	\caption{SNAP board (left) with external AT84AD001B ADC (upper
          right). Initialization and data acquisition is performed
          with a Raspberry Pi (lower right) that connects to the SNAP
          board via a 40-pin GPIO ribbon cable.}
	\label{fig:snap}
\end{figure}

\prizm\ employs two Smart Network ADC Processor
(SNAP\footnote{https://casper.berkeley.edu/wiki/SNAP}) boards to read
out and process the analog RF signals~\citep{Hickish2016}.  Each SNAP
board accepts two RF inputs corresponding to the two polarizations
from a single antenna.  The input signals are sampled at 500~Msamp/s
using a dual, monolithic, eight-bit, AT84AD001B analog-to-digital
converter (ADC).  As shown in Figure~\ref{fig:snap}, the ADC is an
external unit that attaches to the SNAP board via a Z-Dok connector.
The clock signal for the ADC is provided by a Valon~5009 frequency
synthesizer.  The SNAP board is equipped with a
Kintex~7\footnote{http://www.xilinx.com/products/silicon-devices/fpga/kintex-7.html}
field-programmable gate array and is programmed with spectrometer
firmware that uses a four-tap polyphase filter bank to channelize the
two digitized input signals and generate auto- and cross-spectra.  The
spectra span 0--250~MHz with 4096 frequency channels and are recorded
roughly once every four seconds.  A Raspberry Pi (RPi) single-board
computer is used to communicate with the SNAP board, run the data
acquisition software, and store data.  The data rate is sufficiently
low ($\sim$900 MB per day, both antennas combined) that a standard SD
card on the RPi can save several months of data at a time.

\subsection{Auxiliary electronics}

The Faraday cage that encloses the second stage and readout
electronics also houses auxiliary circuitry for controlling the
mechanical switches.  All four switches (one per polarization for both
antennas) are driven by four ULN2003A Darlington transistors operating
in parallel, and the transistors are commanded simultaneously by a
dedicated RPi.  This auxiliary RPi also records the internal
temperature of the Faraday cage with a DS18B20 sensor.  The auxiliary
and two SNAP-control RPis are each equipped with Adafruit Ultimate GPS
Breakout boards.  These boards have integrated real time clocks that
maintain the system times on the RPis, and the clocks are periodically
synchronized to GPS satellites by hand, using an active antenna that
is external to the Faraday cage.

\subsection{Power distribution} \label{sec:power}

The entire \prizm\ system is powered using eight 12-V, 170-Ah Lead
Crystal batteries that are wired in parallel.  The total system power
draw is $\sim$80~W, and \prizm\ can operate without interruption for
roughly one week when the batteries are fully charged.  Battery
charging is performed using a Honda EU30is generator and a fuel cache
that is kept at the \prizm\ observing site.  During observations, the
batteries are connected to a DC/DC converter that is housed inside the
Faraday cage and provides a stable 12~V output despite the slow
decline of the battery voltage.  The 12~V output drives the mechanical
switches, and there are additional regulators that supply lower
voltage levels to various other components in the system.

\begin{figure}
	\centering
	\includegraphics[width=0.6\textwidth]{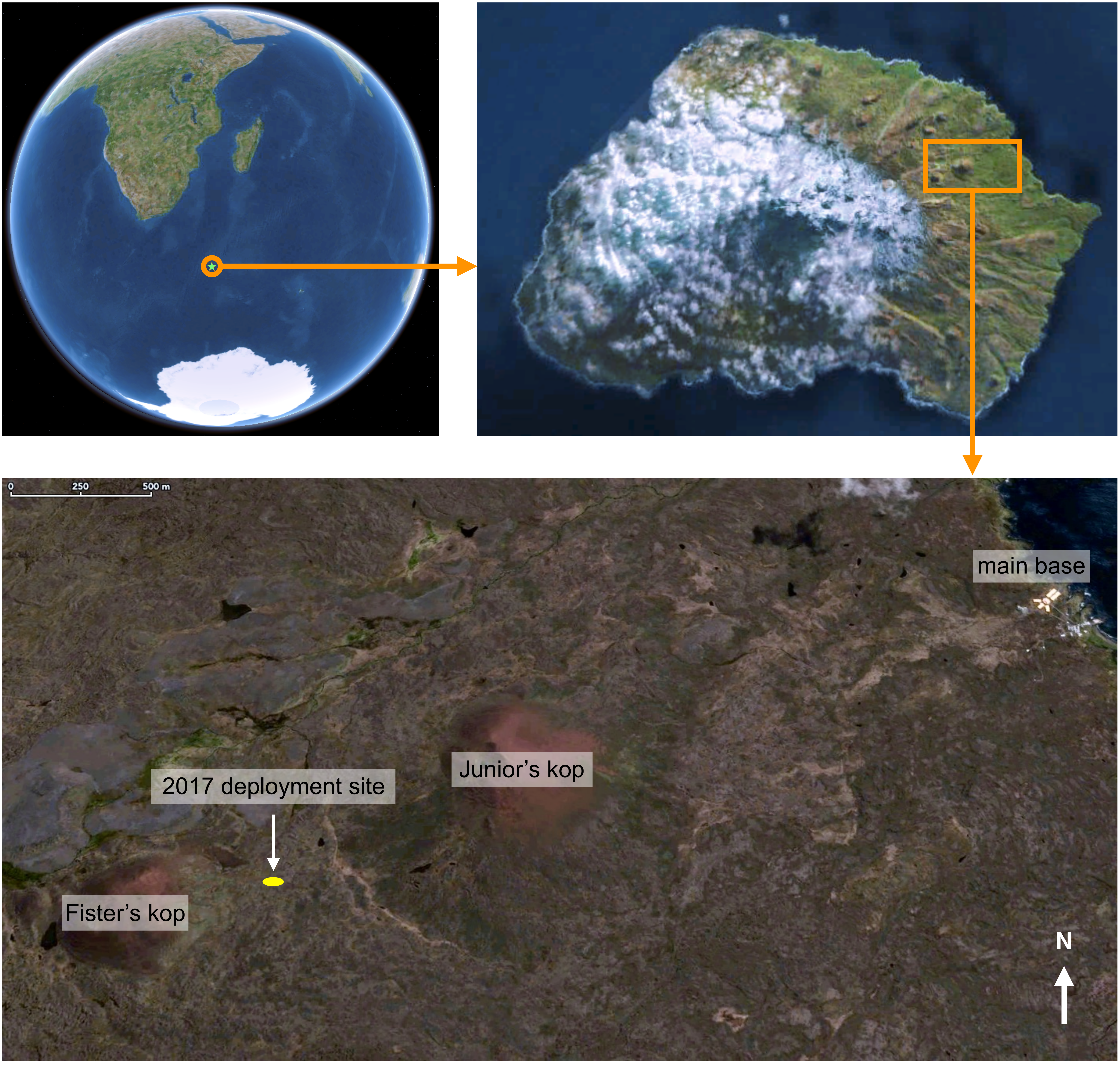}
	\caption{Marion Island is located in the sub-Antarctic,
		roughly halfway between South Africa and Antarctica.  The
		main research base is situated on the northeast side of the
		island, and the \prizm\ observing site is about 4~km inland,
		between Junior's and Fister's kop.}
	\label{fig:base_site}
\end{figure}

\section{Deployment and operations} \label{sec:deployment}

\subsection{Marion Island} \label{sec:marion}

Marion Island (Figure~\ref{fig:base_site}) is a research station that
is operated by the South African National Antarctic Programme and is
located at 46$^\circ$54$'$45$''$S, 37$^\circ$44$'$37$''$E in the
Southern Indian Ocean, about 2000~km away from the nearest continental
land masses. The area of the island is $\sim$290 km$^2$, and the main
research base is situated on the northeast side.  Marion is serviced
via ship once per year in an annual ``relief voyage'' that takes place
in April. The ship visit allows a few-person installation team access
for three weeks. \prizm\ first deployed to Marion Island in April
2017, and the experiment represents the first radio astronomy research
that has been performed from this base.

\begin{figure}
	\centering
	\includegraphics[width=\textwidth]{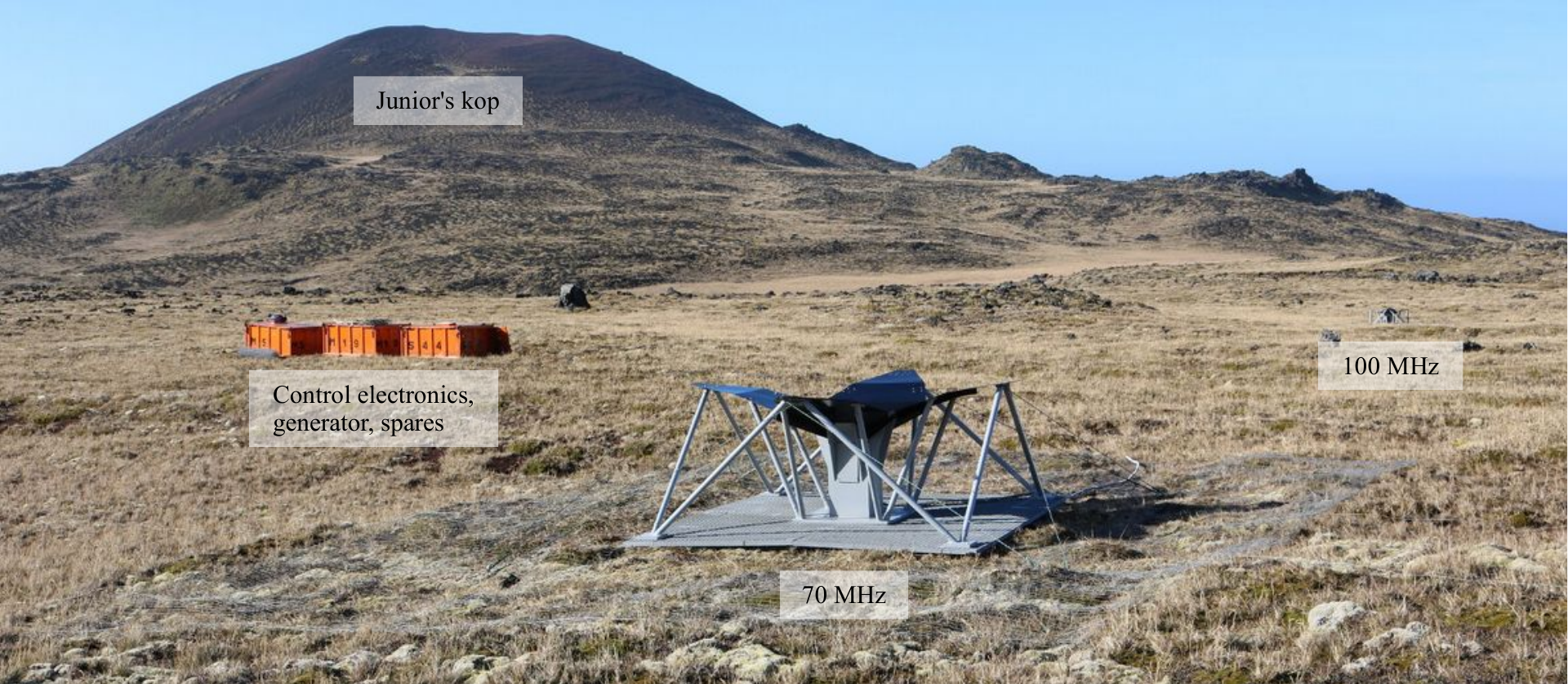}
	\caption{The \prizm\ experiment installed on Marion Island.
		The 70~MHz and 100~MHz antennas are visible in the
		foreground and background, respectively, and the three
		orange shipping containers house the control electronics,
		generator and batteries, fuel cache, and spares.  The main
		base lies 4~km behind Junior's kop.}
	\label{fig:prizm_deployed}
\end{figure}

\subsection{Site selection}

\begin{figure}
	\centering
	\includegraphics[width=0.5\textwidth]{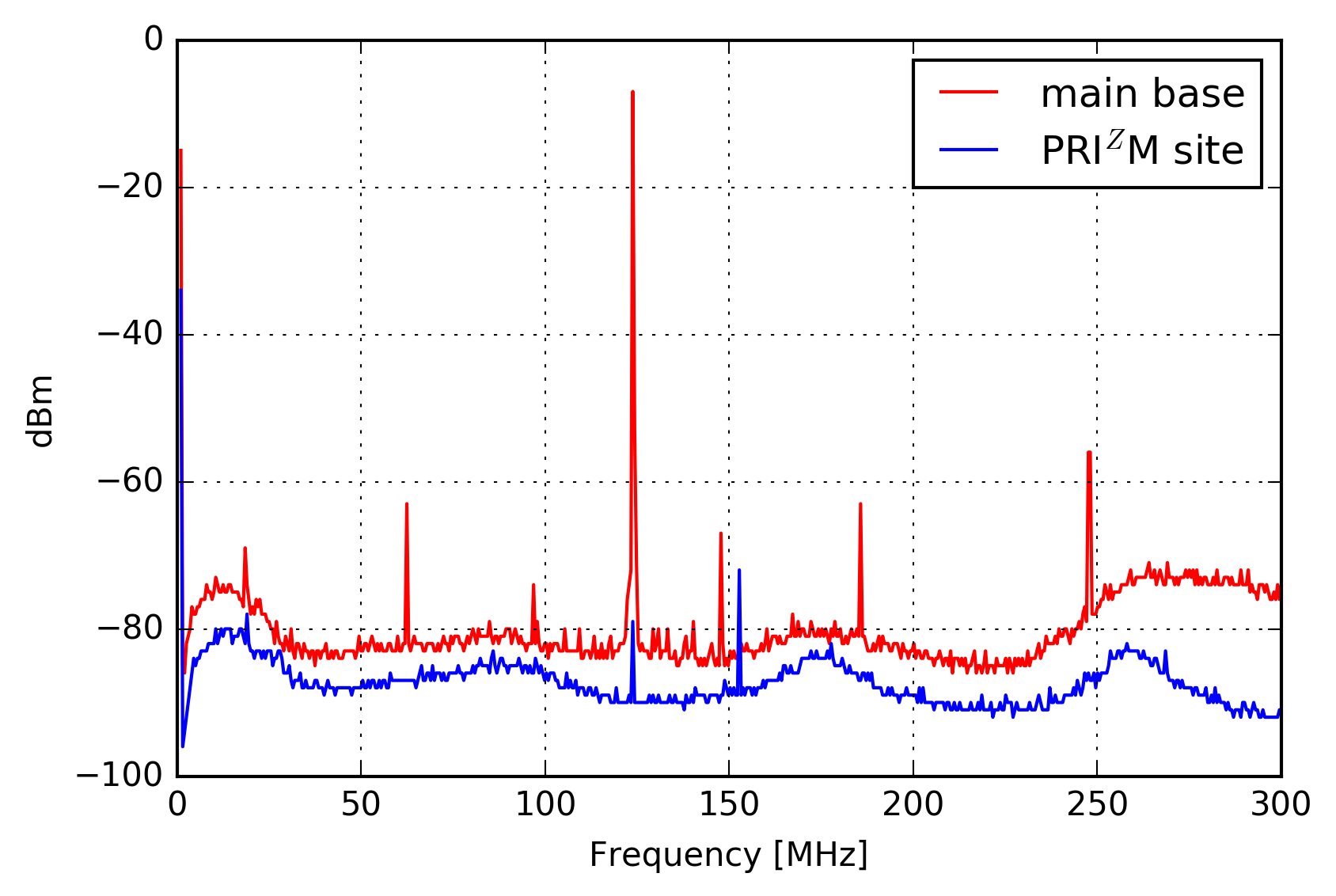}
	\caption{RFI spectrum comparison between the main Marion base
		and the \prizm\ observing location.  These spectra are max
		hold measurements and were taken while a helicopter was
		operating near the base and transmitting at 123.45~MHz.  A
		comparison of the received power at both locations provides
		a rough benchmark of $\sim$60~dB signal suppression, arising
		from a combination of attenuation from Junior's kop and the
		distance between the \prizm\ site and the base.  The peak at
		156~MHz is transmission from a hand held radio.}
	\label{fig:rfi2017}
\end{figure}

The observing site for \prizm\ was selected by balancing several
considerations: ease of regular access from the main base, sufficient
distance from the base to minimize the impact of locally generated
RFI, and suitable terrain.  Junior's kop (see
Figure~\ref{fig:base_site}) is the largest hill within a reasonable
hiking distance from the main base that can provide enhanced RFI
shielding, and we found that the terrain between Junior's and Hendrik
Fister's kop was suitably flat and dry for antenna installation.  We
chose an installation site at 46$^\circ$53$'$13$"$S,
37$^\circ$49$'$10.7$"$E, roughly 4~km southwest from the main base,
and we benchmarked the level of RFI shielding by surveying with a
Workman T-601 discone antenna and a Rohde $\&$ Schwarz spectrum
analyzer.  Figure~\ref{fig:rfi2017} shows a comparative plot of RFI at
the base and the \prizm\ observing site; these measurements were taken
on a day when a helicopter was operating near the base and
transmitting at 123.45~MHz.  The observed power at 123.45~MHz is
suppressed by $\sim$60~dB at the \prizm\ observing site relative to
the base, indicating that any station emissions well be similarly
attenuated. The power reduction arises from a combination of physical
separation from the base and diffraction over Junior's kop.

\subsection{Observations}

\begin{figure}
	\centering
	\includegraphics[width=0.6\textwidth]{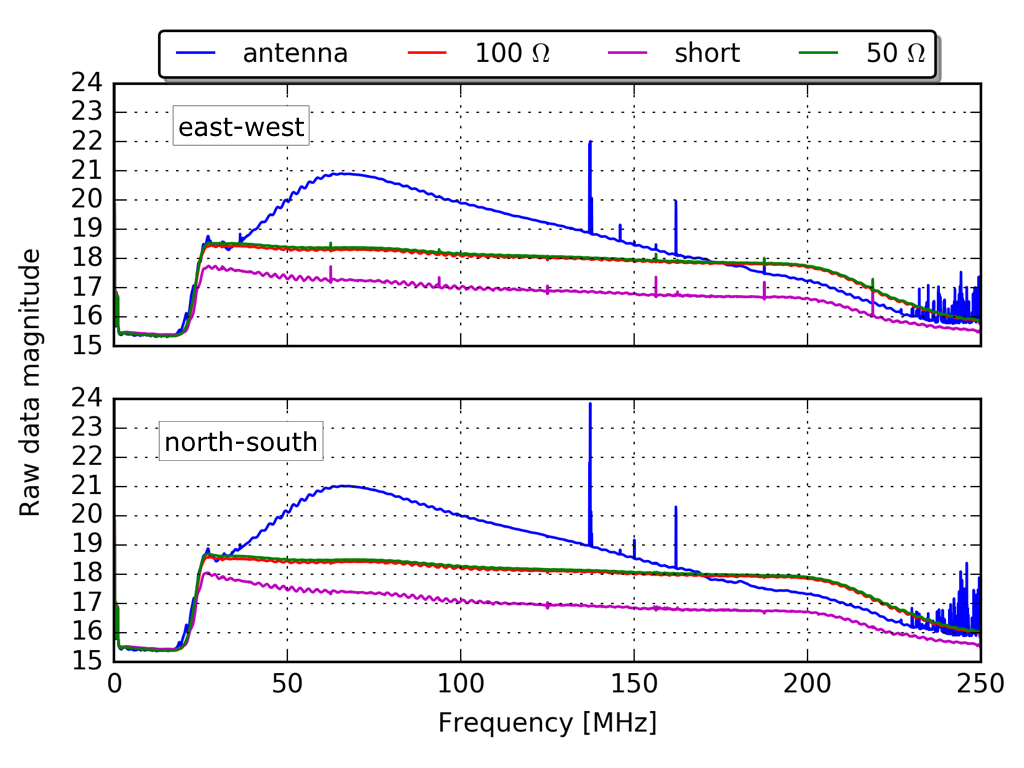}
	\caption{Example uncalibrated auto-spectra from the 100~MHz
		east-west ({\textit{top}}) and north-south
		({\textit{bottom}}) polarizations, taken over a 12-hour
		period.  Averaged spectra are shown for on-sky antenna
		measurements, 100~$\Omega$ and 50~$\Omega$ calibrators, and
		a short.  Transmission from Orbcomm satellites peaks at
		137--138~MHz.}
	\label{fig:data_slice}
\end{figure} 

Figure~\ref{fig:prizm_deployed} shows both \prizm\ antennas deployed
at the observing site.  During normal operations, the mechanical
switches on the antennas follow an observing cadence of one hour on
the sky, followed by one minute on each of the calibrators.  Auto- and
cross-spectra of the north--south and east--west polarizations are
accumulated and written every $\sim4$~seconds for each antenna.

\begin{figure}
	\centering
	\includegraphics[width=0.75\textwidth]{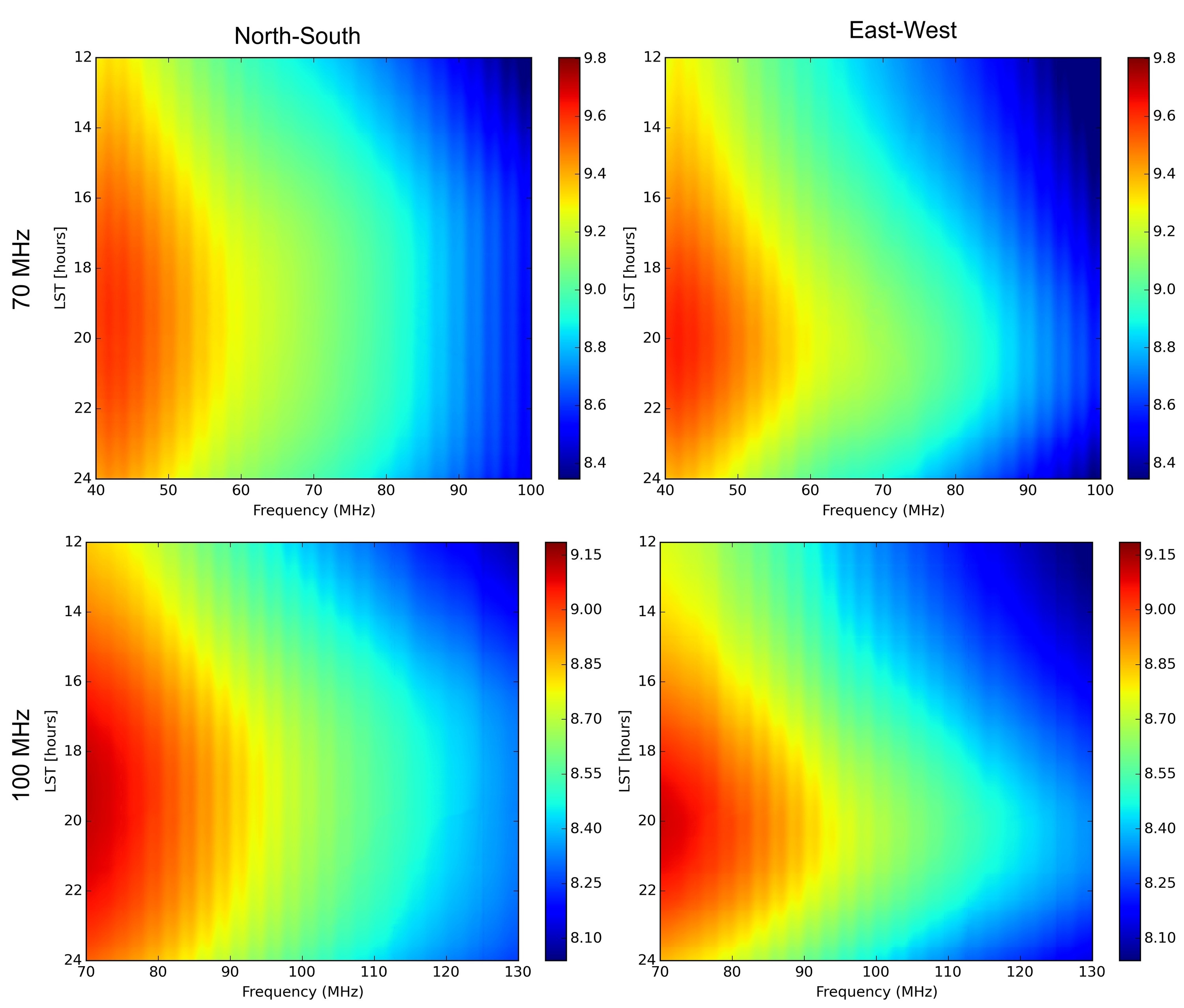}
	\caption{On-sky spectra for both polarizations on the 70 and
		100~MHz antennas, taken over a 12~hour period.  The data are
		uncalibrated, and the color bar is logarithmic in raw ADC
		units.}
	\label{fig:data_waterfall}
\end{figure}

Figure~\ref{fig:data_slice} shows representative raw, averaged spectra
from the 100~MHz antenna observing the sky and the calibrators.  The
falling spectrum of the Galaxy is clearly visible in the on-sky
antenna data, and apart from Orbcomm transmission (137--138~MHz) and
clock harmonics, there is no visible RFI contamination within the
30--200~MHz range.  The noise power spectra of the 50~$\Omega$ and
100~$\Omega$ resistors have similar amplitudes ($N_p$) because they
are at the same physical temperature ($T$).  The nominal relationship $N_p = kTB$ is independent of resistance and depends only on the
bandwidth $B$ and Boltzmann's constant $k$.  This equation assumes a
perfect match between the emitting resistor and the LNA; we use the
two different resistors to quantify any mismatch and to separate
voltage and current noise in the LNA model.

\begin{figure}
	\centering
        \includegraphics[width=0.5\textwidth]{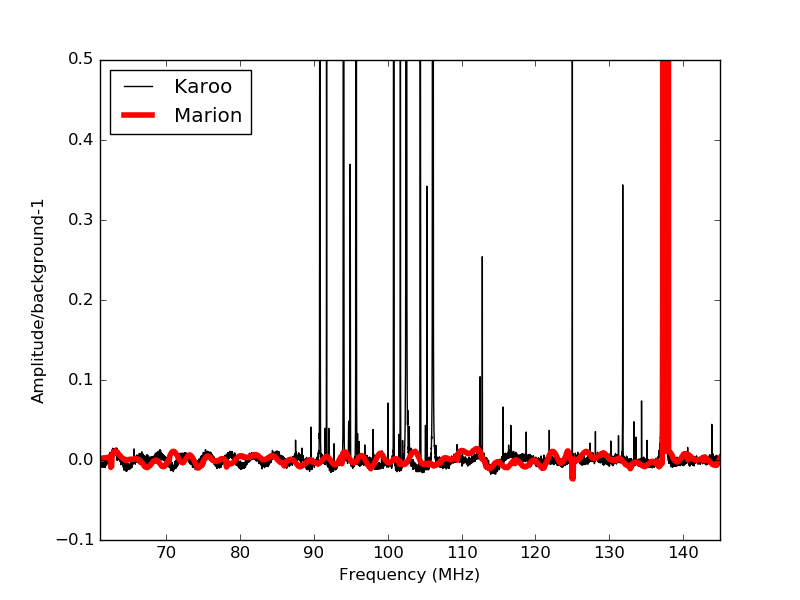}
	\caption{Comparison of radio spectrum on Marion (thick red) and in the
		Karoo desert (thin black) in the cosmic dawn frequency range band from the 100 MHz
                antennas. This plot shows 
                the fractional amplitude above a background fit to the raw and
		uncalibrated data without RFI excision.  Apart
		from Orbcomm satellite transmission at 137--138~MHz, there
		is no visible RFI in the Marion data. The feature at 125 MHz is an artifact
                of the 250 MHz FPGA clock and not due to RFI.}
	\label{fig:marion_karoo}
\end{figure} 

Figure~\ref{fig:data_waterfall} illustrates uncalibrated, on-sky
spectra from both polarizations of the 70 and 100~MHz antennas, taken
over a 12-hour period.  Apart from removing the regular hourly
calibrator observations, there have been no other cuts or processing
applied to the data.  The large-scale spectral variations arise from
the Galaxy drifting through the beam as the Earth rotates.  Averaging
the 100~MHz east-west polarization data within this time period and
applying a high-pass filter yields the spectrum shown in
Figure~\ref{fig:marion_karoo}.  The figure also shows an identically
filtered \scihi\ spectrum taken from the South African Square
Kilometre Array (SKA) site in the Karoo desert.  The filtering
highlights RFI line features, and the spectral comparison illustrates
that the cleanliness of the Marion Island RFI environment far exceeds
the Karoo, with essentially no visible contamination in the FM band.

\section{Discussion and future work}

The installation of the \prizm\ antennas on Marion Island in 2017
represents the first radio astronomy research conducted from this
research base.  The remote location of Marion presents a number of
challenges, including the short three-week access window that is
available only once per year, the harsh environmental conditions of
the Roaring Forties (high winds, rain, cold temperatures), and mouse
infestation.  We have successfully overcome these challenges, and we
have several improvements planned for the \prizm\ system that will be
implemented during subsequent voyages and described in future
publications.

We have found that the remoteness of Marion Island results in an
exceptionally radio-quiet environment, with no evidence of detected
RFI contamination in the FM band.  In particular, the $\sim$2000~km
separation between the island and the mainland is close to the
distance limit at which meteor trails effectively scatter
RFI~\citep[{\textit{e.g.}}][]{Wislez96}.  We have not seen
qualitative evidence of enhanced RFI from meteor scattering, an effect
that is commonly visible at many other remote sites.  The overall
radio quietness of Marion Island surpasses even the South African SKA
site in the Karoo desert.

The lack of RFI, in combination with calmer ionospheric conditions
during winter nights, make Marion an excellent site for low-frequency
radio observations.  The International Reference Ionosphere 2012
model\footnote{https://iri.gsfc.nasa.gov/} suggests that during the
last solar minimum, the ionosphere plasma frequency dropped as low as
1.5~MHz on winter nights.  The \prizm\ site will provide the necessary
infrastructure and test bed to deploy new low-frequency experiments in
the future, possibly opening new observational windows at long wavelengths
that have been prohibitively difficult to measure to date.

\section*{Acknowledgments}

We gratefully acknowledge the National Research Foundation (grant
number 93087) and the South African National Antarctic Programme for
providing funding and logistical support for our research program.
The financial assistance of the South African SKA Project (SKA SA)
towards this research is hereby acknowledged (www.ska.ac.za).  We
additionally extend our sincere gratitude to the staff at SKA SA for
hosting our lab tests.  We thank the South African National Space
Agency for their technical support and the crew of Ultimate Heli for
safely delivering us and our cargo.  The authors also wish to thank
Judd Bowman, Matt Dexter, Ryan Monroe, Raul Monsalve, Aaron Parsons
and Nipanjana Patra for useful discussions.

\bibliographystyle{ws-jai}
\bibliography{Bibliography}{}
\end{document}